\documentclass[twocolumn,superscriptaddress,showpacs,preprintnumbers,amsmath,amssymb,prb]{revtex4}
\newcommand{\graphwidth}{0.9\columnwidth}

\usepackage{graphicx}
\usepackage{dcolumn}
\usepackage{bm}
\usepackage{color}
\usepackage[hypertex,colorlinks=true,linkcolor=black,citecolor=blue,filecolor=blue,urlcolor=blue,setpagesize=false]{hyperref}

\newcommand{\msr}{$\mu$SR}
\newcommand{\LFAOF}{LaFeAsO$_{1-x}$F$_x$}
\newcommand{\CFCAF}{CaFe$_{1-x}$Co$_x$AsF}
\newcommand{\BFCA}{Ba(Fe$_{1-x}$Co$_x)_2$As$_2$}
\newcommand{\BKFA}{Ba$_{1-x}$K$_x$Fe$_2$As$_2$}
\newcommand{\PRL}{Phys.~Rev. Lett. }
\newcommand{\PRB}{Phys.~Rev. B }
\newcommand{\JPSJ}{J.~Phys. Soc. Jpn. }
\newcommand{\etal}{{\it et}~{\it al}.}

\begin{document}
\preprint{APS/123-QED}
\title{Cooperative order in the weakly magnetic domain of LaFeAsO$_{1-x}$F$_{x}$ near the doping phase boundary}
\author{Masatoshi Hiraishi}
\affiliation{Muon Science Laboratory and Condensed Matter Research Center, Institute of Materials Structure Science, High Energy Accelerator Research Organization (KEK), Tsukuba, Ibaraki 305-0801, Japan}
\author{Ryosuke Kadono}
\affiliation{Muon Science Laboratory and Condensed Matter Research Center, Institute of Materials Structure Science, High Energy Accelerator Research Organization (KEK), Tsukuba, Ibaraki 305-0801, Japan}
\affiliation{Department of Materials Structure Science, The Graduate University for Advanced Studies, Tsukuba, Ibaraki 305-0801, Japan}
\author{Masanori Miyazaki}
\affiliation{Muon Science Laboratory and Condensed Matter Research Center, Institute of Materials Structure Science, High Energy Accelerator Research Organization (KEK), Tsukuba, Ibaraki 305-0801, Japan}
\author{Ichihiro Yamauchi}
\affiliation{Muon Science Laboratory and Condensed Matter Research Center, Institute of Materials Structure Science, High Energy Accelerator Research Organization (KEK), Tsukuba, Ibaraki 305-0801, Japan}
\author{Akihiro Koda}
\affiliation{Muon Science Laboratory and Condensed Matter Research Center, Institute of Materials Structure Science, High Energy Accelerator Research Organization (KEK), Tsukuba, Ibaraki 305-0801, Japan}
\affiliation{Department of Materials Structure Science, The Graduate University for Advanced Studies, Tsukuba, Ibaraki 305-0801, Japan}
\author{Kenji M.~Kojima}
\affiliation{Muon Science Laboratory and Condensed Matter Research Center, Institute of Materials Structure Science, High Energy Accelerator Research Organization (KEK), Tsukuba, Ibaraki 305-0801, Japan}
\affiliation{Department of Materials Structure Science, The Graduate University for Advanced Studies, Tsukuba, Ibaraki 305-0801, Japan}
\author{Motoyuki Ishikado}\thanks{Present address: Research Center for Neutron Science and Technology, Comprehensive Research Organization for Science and Society (CROSS), Tokai-mura, Naka-gun, Ibaraki 319-1106, Japan}
\affiliation{Japan Atomic Energy Agency, Muramatsu, Tokai-mura, Naka-gun, Ibaraki 319-1184, Japan}
\author{Shuichi Wakimoto}
\affiliation{Japan Atomic Energy Agency, Muramatsu, Tokai-mura, Naka-gun, Ibaraki 319-1184, Japan}
\author{Shin-ichi Shamoto}
\affiliation{Japan Atomic Energy Agency, Muramatsu, Tokai-mura, Naka-gun, Ibaraki 319-1184, Japan}

\date{\today}
\begin{abstract}
Spatial phase separation into mesoscopic domains of non-magnetic [superconducting (SC) below $T_{\rm c}\simeq25.5$~K] and two kinds of magnetic phases, one showing disordered spin density wave (d-SDW) order and another associated with glassy weak magnetism (WM), are observed below $\sim$100~K by muon spin rotation (\msr) in \LFAOF\  for $x=0.057(3)$ which is near the boundary of these phases on the doping phase diagram. In contrast to the competing order observed in the regular SDW phase of \BFCA, the WM domain exhibits cooperative coupling of superconducting and magnetic order parameters as inferred from strong diamagnetism and associated negative shift of \msr\ frequency just below $T_{\rm c}$.
\end{abstract}
\maketitle
Since the discovery of high-$T_{\rm c}$ superconductivity in iron-based compounds\cite{Kamihara}, the interplay between magnetism and superconductivity has been one of the most fascinating topics because they comprise ``magnetic" ions that hinder superconductivity in conventional superconductors. Interestingly, there are a couple of reports suggesting spatially homogeneous (microscopic) coexistence of superconducting (SC) and magnetic (spin density wave, SDW) phases in \BFCA\ near the boundary of these two phases, where clear competition between the SC and SDW order parameters is observed, e.g., reduction of the iron moments\cite{Pratt} and/or orthorhombic distortion parameter $\delta\equiv(a-b)/(a+b)$ under strong magneto-elastic coupling\cite{Nandi} just below superconducting transition temperature ($T_{\rm c}$).

Another mode of coexistence is the segregation of these two phases into {\sl mesoscopic} domains. Such a symbiotic situation has been suggested for underdoped \BKFA\
from earlier studies using local probes including muon spin rotation (\msr)\cite{Aczel:08,Goko:09,Park:09} and NMR\cite{Julien:09}, while there also exist number of reports arguing for microscopic coexistence\cite{Rotter:08,Chen:09,Rotter:09,Wiesenmayer}. A similar mesoscopic phase separation is also observed in \CFCAF, where the magnetic domain turns out to exhibit highly disordered magnetism distinctly different from regular SDW phase of pristine compound\cite{Takeshita_CFCAF}.

It should be stressed that there is no dichotomy between these two modes of ``coexistence," considering possible microscopic coexistence of these two phases {\sl within} the segregated magnetic domains. Here, we report \msr\ measurements in the prototype compound LaFeAsO$_{1-x}$F$_{x}$ (LFAOF) that demonstrates yet another type of correlation between magnetism and superconductivity near the doping phase boundary.
It is inferred in a sample with $x=0.057(3)$ that spatial segregation of electronic ground state into non-magnetic (SC below $T_c$) and two different kinds of magnetic phases occurs below $\sim$100~K. One of these magnetic phases is disordered spin density wave phase (d-SDW, similar to the corresponding phase in \CFCAF) and the other is characterized by spin glass-like weak magnetism (WM). The domain of the WM phase, which does not accompany orthorhombic distortion, exhibits anomalous diamagnetism below $T_{\rm c}=25.5$~K, where the demagnetization is much greater than that due to formation of flux line lattice (FLL) in the SC domain. These observations are perfectly in line with our previous report on a different LFAOF sample ($x\simeq0.06$)\cite{Takeshita_LFAOF}. The enhanced diamagnetism in the WM domain that is presumably situated in the proximity of SC domain suggests a novel mode of cooperative coupling between these two order parameters in the WM domain.

Despite the known difficulty in the synthesis of single crystal that precludes detailed study of LFAOF and so-called 1111-family compared with other iron-based superconductors\cite{Prozorov, Zhigadlo, Jia, Ishikado_single}, LFAOF is still favorable for investigating a relationship between magnetism and superconductivity using sensitive magnetic probe like \msr\ because it is free from magnetic rare-earth ions that tends to mask magnetism arising from Fe$_2$As$_2$ planes.

Conventional \msr\ experiments were performed using D$\Omega$1 spectrometer installed on the D1 beamline at J-PARC MUSE Facility, Japan, and Hi-Time spectrometer on the M15 beamline at TRIUMF, Canada. A 100\% spin-polarized beam of positive muons with a momentum of 27 MeV/c was irradiated to a polycrystalline sample of LFAOF loaded to a He-flow cryostat to monitor positron decay asymmetry $A(t)$.
During the measurement under a zero field (ZF), residual magnetic field at the sample position was reduced below 10$^{-6}$~T with the initial muon spin direction parallel to the muon beam direction. All the measurements under a magnetic field were performed by cooling sample to the target temperature after the field equilibrated to eliminate the effect of flux pinning.
Florine concentration of the sample was determined to be $x=0.057(3)$ by secondary ion mass spectrometry (SIMS). Details of sample preparation are described in the earlier reports~\cite{Ishikado, Wakimoto}.

\begin{figure}[!t]
 \centering
  \includegraphics[width=\graphwidth,clip]{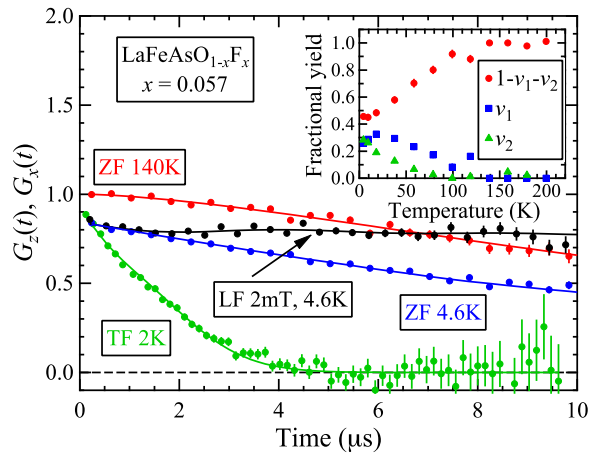}
  \caption{(Color online). \msr\ time spectra under various field conditions. Only envelope curve is displayed for the spectrum under a transverse field (TF=0.05~T, observed at 2~K). Inset: volumetric fraction of non-magnetic ($1-v_1-v_2$) and magnetic phases ($v_1$ for WM, $v_2$ for d-SDW) determined by ZF-\msr\ as a function of temperature.}
  \label{Fig_Spectra1}
\end{figure}

Figure \ref{Fig_Spectra1} shows typical examples of normalized \msr\ time spectra $G_\alpha(t)=A(t)/A_0$ [where $\alpha=z$ for ZF/LF and $=x$ for TF, and $A_0=A(0)$] under various conditions of external field. ZF-\msr\ spectrum at 140~K shows a Gaussian-like slow depolarization due to weak random local fields from nuclear magnetic moments which is described by the Kubo-Toyabe function $G^{\rm KT}_z(t)$ ($\simeq e^{-\sigma_{\rm n}^2t^2}$ for $\sigma_{\rm n}t\ll1$), indicating that the entire sample is non-magnetic.
Meanwhile, ZF-\msr\ spectra exhibit apparent reduction of the initial asymmetry with decreasing temperature [i.e., $G_z(0)<1$ as $A_0$ has been fixed to the value at ambient temperature in extracting $G_z(t)$ in Fig.~\ref{Fig_Spectra1}]. Additional ZF-\msr\ measurement with higher time resolution ($\sim$1 ns) reveals that this is due to a fast depolarization ($\Lambda_2\simeq20$ MHz) without oscillation. Considering that the magnitude of internal field estimated from $\Lambda_2$ is comparable to that in the SDW phase at 2~K in the pristine compound ($x=0$), this rapid depolarization is attributed to the onset of highly inhomogeneous internal field associated with the development of the d-SDW phase.

It is also noticeable that the line shape of the component showing slow depolarization changes from Gaussian-like to exponential with decreasing temperature (see 140~K versus 4.6~K in Fig.~\ref{Fig_Spectra1}).
We attributed this to development of another magnetic phase characterized by spin glass-like weak magnetism (WM).
The slow exponential depolarization is quenched by applying a weak longitudinal field (LF, parallel to the initial muon spin polarization) at the lowest temperature, indicating that the depolarization is due to highly inhomogeneous internal field distribution that is quasi-static within the time window of \msr.
Considering these, we used the following form for the curve-fit analysis of the ZF-\msr\ time spectra;
\begin{align}
  A(t)&=A_0G^{\rm KT}_z(t)\left[(1-v_1-v_2)+{\scriptstyle \sum_{i=1,2}}v_i g_i(t)\right],\\
	g_i(t)&\equiv \frac{1}{3}+\frac{2}{3}(1-\Lambda_i t)e^{-\Lambda_i t}
\end{align}
where $v_i$ and $\Lambda_i$ are the fractional yield and depolarization rate of the WM ($i=1$) and d-SDW ($i=2$) domains, respectively. 
The residual term (1/3) in $g_i(t)$ represents the probability for muons implanted to the magnetic domains to be exposed to internal field $B$ parallel with the initial muon polarization (i.e., $\int\cos^2\theta d\cos\theta=1/3$, with $\theta$ being the angle between $B$ and $z$).
We note that the spectra obtained at J-PARC were analyzed by eqs.~(1) and (2) with $g_2(t)\simeq1/3$ because of the large $\Lambda_2$ compared with the instrumental time resolution ($\sim$100 ns). 
Temperature dependence of $v_i$ is shown in the inset of Fig.~\ref{Fig_Spectra1}, where the magnetic phases emerge below $\sim$100~K, and gradually develops with decreasing temperature to reach $v_1\sim0.33(1)$ and $v_2\sim0.20(1)$ at the lowest temperature. Considering that this sample is single-phased as inferred from powder x-ray diffraction data~\cite{Wakimoto}, we conclude that the spatial segregation of magnetic phases is intrinsic. 

Applying a transverse field (TF, perpendicular to the initial muon polarization) $B_0$ to sample in the SC state brings inhomogeneity to the internal field distribution due to the formation of FLL associated with superconductivity, which is observed as an enhanced depolarization in TF spectra, as shown in Fig.~\ref{Fig_Spectra1}.
It is reasonably presumed that implanted muons are distributed randomly over the length scale of FLL, probing local magnetic fields at their respective positions.
Then TF-\msr\ signal consists of a random sampling of internal field distribution $B({\bf r})$, such that
	$P^{\rm v}_x(t)\simeq\int_{-\infty}^{\infty}\cos\left(\gamma_{\mu}Bt+\phi\right)n(B)dB,$
where $\gamma_{\mu}=2\pi\times135.53$ MHz/T is the muon gyromagnetic ratio, $n(B)$ is the density distribution of $B$ over the entire sample volume, and $\phi$ is the initial phase of rotation. Hence, $n(B)$ can be deduced from the Fourier transform of the TF-\msr\ spectra (although one must consider the limited time window of observation due to muon decay lifetime $\sim$10$^1$ $\mu$s).
In the case of relatively long magnetic penetration depth ($\geq300$~nm) and/or of polycrystalline samples, Gaussian distribution is a reasonable approximation for $n(B)$, yielding
\begin{align}
	P^{\rm v}_x(t)\simeq G_x(t)\cos\left(\omega_{\rm s} t+\phi\right)\simeq e^{-\sigma_{\rm s}^2t^2/2}\cos\left(\omega_{\rm s} t+\phi\right),
\end{align}
where $\sigma_{\rm s}$ is obtained from a second moment of the field distribution ($=\gamma_{\mu}\sqrt{\langle[B({\bf r})-B_0]^2\rangle}$), and $\omega_{\rm s}\simeq\gamma_{\mu}B_0$.
The complete depolarization of TF spectrum in Fig.~\ref{Fig_Spectra1} $[G_x(t)\rightarrow0$ with $t\rightarrow\infty]$ indicates that the entire volume of non-magnetic phase falls into the SC (FLL) state.

\begin{figure}
 \centering
  \includegraphics[width=\linewidth,clip]{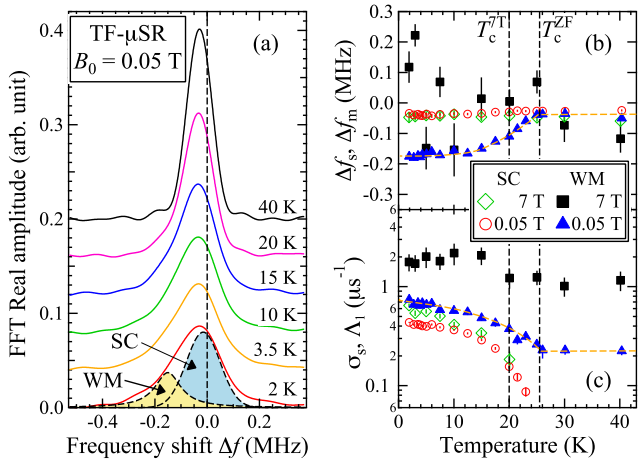}
  \caption{(Color online). (a) Fast Fourier transform (real amplitude) of TF-\msr\ time spectra measured at 0.05~T, where the horizontal axis is the shift from the central frequency. Yellow (blue) hatched area shows signal from WM (SC) domain. Spectra above 3.5~K are shifted upwards for visibility. (b), (c) Temperature dependence of frequency shift and relaxation rate at $B_0=0.05$ and 7 T deduced for the respective domains. Dashed curves/lines in (b) and (c) are guides to the eye, where $T_{\rm c}$ at 7~T is quoted from Ref.~\onlinecite{Shahbazi}.}
  \label{Fig_TF}
\end{figure}

Fast Fourier transform (FFT) of TF-\msr\ time spectra measured at 0.05~T are shown in Fig.~\ref{Fig_TF}~(a), where a reference frequency ($\simeq\gamma_\mu B_0$) for the shift is defined by the signal from muons stopped in a plastic scintillator positioned right behind the sample.
The FFT spectra show a symmetric profile above $T_{\rm c}$ except a small negative shift, indicating that $B\simeq B_0$.  They become broad on cooling below $T_{\rm c}$, which at a glance seems to simulate the effect of FLL. However, it is noticeable that the enhancement of amplitude occurs at lower fields, which cannot be explained by $n(B)$ associated with the FLL state. Considering the result of ZF-\msr, we attribute this low field tail to the signal from muons stopped in the WM domain, and
 we analyzed TF-\msr\ spectra in time domain using the following form, 
\begin{align}
	A(t)\simeq&A'_0e^{-\sigma_{\rm n}^2t^2/2}\Bigl[v_1e^{-\Lambda_1 t/\sqrt{2}}\cos(\omega_{\rm m}t+\phi)\nonumber\\
	&+(1-v_1)P^{\rm v}_x(t)\Bigr],\label{tfsp}
\end{align}
where $\sigma_{\rm n}$ denotes the depolarization rate due to random local fields from nuclear magnetic moments, $\Lambda_1$ and $\omega_{\rm m}$ denote the depolarization rate and precession frequency in the WM domain, and $v_1$ is fixed to the value obtained by curve fits of ZF-\msr\ at the corresponding temperature. Note that signal originating from d-SDW domain is invisible because $\Lambda_2\sim$20~MHz [i.e., $A'_0\simeq A_0(1-v_2)$]. In addition, $\sigma_{\rm n}$ can be assumed to be independent of temperature, and thereby it is fixed to the value at 40~K [above $T_{\rm c}$, $\sigma_{\rm n}=0.119(4)$ $\mu$s$^{-1}$] for the curve fits of the spectra using eq.~(\ref{tfsp}).

The result of analysis for the data obtained for $B_0=0.05$ and 7~T is summarized in Fig.~\ref{Fig_TF}, where the shift of precession frequency from the value at 40~K, $\Delta f_{\rm s,m}(T)=[\omega_{\rm s,m}(T)-\omega_{\rm s,m}(40\:{\rm K})]/2\pi$, is plotted together with the depolarization rates deduced from the curve fit.  Although it is not obvious in Fig.~\ref{Fig_TF}(b), $\Delta f_{\rm s}$ exhibits slight decrease below $T_{\rm c}$. This, together with the increase of $\sigma_{\rm s}$ shown in Fig.~\ref{Fig_TF}(c) clearly indicates the formation of FLL state in the SC domain, where the onset temperature of increase is perfectly in line with $T_{\rm c}$ at respective fields determined by bulk properties\cite{Shahbazi}.

Surprisingly, $\Delta f_{\rm m}$ in the WM domain also exhibits a decrease just below $T_{\rm c}$, which disappears at 7~T (or even showing a tendency of slight increase with decreasing $T$). The relative magnitude of shift is remarkably large as  $\Delta f_{\rm m}/f_{\rm m}\simeq-1$\% at 0.05 T that far exceeds the change induced by FLL formation in the SC domain ($\Delta f_{\rm s}$).  Moreover, $\Lambda_1$ also exhibits an increase below $T_{\rm c}$, indicating that inhomogeneity of field distribution in the WM domain is enhanced just below $T_{\rm c}$.  These anomalous behaviors observed in the WM phase are perfectly in accordance with our earlier report on a different sample ($T_{\rm c}\simeq18$~K)~\cite{Takeshita_LFAOF}, confirming that the observed phenomenon under a transverse field is intrinsic in \LFAOF.


As mentioned above, the magnitude of $\Delta f_{\rm m}$ [$=-0.136(6)$~MHz $\gg \Delta f_{\rm s}=-0.011(4)$~MHz] at 0.05~T indicates that the observed shift in the WM domain cannot be explained by the FLL formation.  The magnitude of $\Delta f_{\rm s}$ is estimated by calculating $n(B)$ for the known values of magnetic field penetration depth [$=495(3)$ nm at 2 K, deduced from $\sigma_{\rm s}$] and Ginzburg-Landau coherence length ($\sqrt{\Phi_0/2\pi H_{\rm c2}}\simeq2.56$~nm with $\mu_0H_{\rm c2}\simeq50$~T~\cite{Hunte}) to yield $\simeq-0.042$~MHz,  which is consistent with the present result.  Thus, it is concluded that the large negative shift in the WM domain just below $T_{\rm c}$ is not due to the flow of orbital currents associated with the conventional proximity effect.

In the case of Ba-122 compounds where the microscopic coexistence is observed, the iron moment size of SDW phase is gradually reduced below $T_{\rm c}$ without change in the moment direction or local symmetry\cite{Pratt,Wiesenmayer}. The phenomenon seems to accompany suppression of orthorhombic lattice distortion as well\cite{Wiesenmayer,Nandi}.
A similar behavior is reported by NMR measurement on BaFe$_2$(As$_{1-x}$P$_x$)$_2$\cite{Iye}, which is understood by considering a quadratic term, $\frac{\gamma}{2}M^2|\Psi|^2$ ($\gamma>0$), for the coupling of order parameters between SDW ($M$) and SC ($\Psi$) in a Landau expansion of free energy\cite{Fernandes:10} under strong magneto-elastic coupling.

The negative shift in LFAOF, however, cannot be explained by the reduction of staggered iron moments and associated change of internal field at muon site; if so, the mean value of the random local fields in the WM domain would have remained unchanged (i.e., $\Delta f_{\rm m}=0$) while $\Lambda_1$ would have been reduced [which is actually observed by NMR in BaFe$_2$(As$_{1-x}$P$_x$)$_2$\cite{Iye}], where the latter is predicted from the equation
\begin{equation}
\frac{\Lambda_1^2}{2}=\gamma_\mu^2\sum_i(m_iA^z_i)^2=\gamma_\mu^2\sum_{i,\alpha}\frac{m_i^2}{r_i^6}
\left(\delta_{\alpha z}-\frac{3r_i^\alpha r_i^z}{r_i^2}\right)^2,\label{lnwth}
\end{equation}
where $m_i$ is the moment size of $i$-th iron ion at a distance $r_i$ from muon, $A^z_i$ is the dipole tensor, $\alpha=x,y,z$, and $\delta_{\alpha z}$ is the Kronecker's delta. Thus, the observed phenomenon in the WM domain is different from that in the regular SDW phase of Ba-122 family.

The WM domain is distinct from regular SDW phase by the absence of orthorhombic distortion \cite{Qureshi:10} ($\delta\simeq0$, which is common to the case in CaFe$_{1-x}$Co$_x$AsF\cite{Nomura:09}), suggesting that the origin of disorder is the frustration of electronic states associated with strong magneto-elastic coupling. This leads us to speculate a novel situation that the term $\frac{\gamma}{2}M^2|\Psi|^2$ becomes negative in the WM domain (i.e., $\gamma\equiv-\gamma_\chi<0$), so that the relevant interaction with proximity to SC phase would lift the degeneracy to release the excess magnetic entropy.

\begin{figure}[t]
 \centering
  \includegraphics[width=\graphwidth,clip]{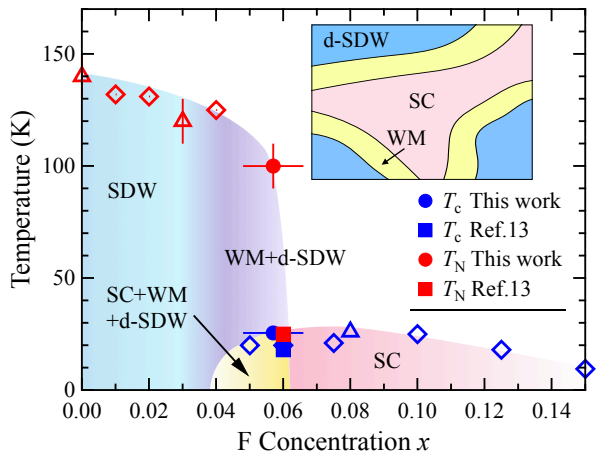}
  \caption{(Color online). Electronic phase diagram of superconducting transition temperature $T_{\rm c}$ (blue) and magnetic transition temperature (red) in \LFAOF. Superconductivity and magnetism are ``symbiotic" in yellow region where WM phase is observed. Data represented by open symbols are quoted from Refs.\onlinecite{Wakimoto, Takeshita_LFAOF, Luetkens, Carlo}. Inset: a schematic drawing of mesoscopic phase separation into d-SDW, WM, and non-magnetic (SC) domains for $x=0.057$.}
  \label{Fig_Phs_dgrm}
\end{figure}

In the conventional proximity effect, the SC order parameter $\Psi$ permeates into the normal state domain as a function of distance $d$ from the boundary,
\begin{equation}
\Psi_{\rm N}\simeq c\Psi\exp(- |d|/\xi_{\rm N})\label{psin},
\end{equation}
where $c$ is a constant ($<1$) and $\xi_{\rm N}$ is the coherence length in the normal state. Note that the Cooper pairing is gapless in the normal state domain, while $|\Psi_{\rm N}|>0$. $\xi_{\rm N}$ is determined by the Fermi velocity in the clean limit, and close to $10^1$-$10^2$ nm over the relevant temperature range in LFAOF, which hints the length scale of the WM domain.
Then the coupling $-\frac{\gamma_\chi}{2}M^2|\Psi_{\rm N}|^2$ induces additional development of magnetic order parameter for $T<T_c$ with diamagnetic tendency to preserve $|\Psi_{\rm N}|$, which would be in proportion to the quantity
\begin{equation}
|\Psi_{\rm N}|\cdot M=|\Psi_{\rm N}|\cdot\sum_i\Delta m_i\:, \label{nop}
\end{equation}
where $\sum_i\Delta m_i\equiv\sum_i\{m^z_i(H)-m^z_i(0)\}\propto\chi H$.
This is probed as a Knight shift via hyperfine coupling of muon to the staggered moment,
\begin{equation}
\Delta f_{\rm m}=|\Psi_{\rm N}|\cdot\frac{\gamma_\mu}{2\pi} \sum_i\Delta m^z_i\cdot A^z_i\propto|\Psi_{\rm N}|\chi H.\label{dFm}
\end{equation}
The enhancement of $\Lambda_1$ is naturally explained as an rms of eq.~(\ref{dFm}).
It is clear from eq.~(\ref{psin}) that the magnitude of the relevant effect depends on the domain size $\sim d$ in relation to $\xi_{\rm N}$, and $d$ might have been too large to induce susceptible effect in CaFe$_{1-x}$Co$_x$AsF where the d-SDW phase predominates over a wide range of $x$\cite{Takeshita_CFCAF}.
At this stage, it would be interesting to note that the above mentioned phenomenon is remarkably similar to that observed in the so-called $A$ phase of a prototype heavy fermion superconductor, CeCu$_2$Si$_2$ (Ref.\onlinecite{Koda:02}), where the $A$ phase develops near the boundary of SDW and SC phases and exhibits random magnetism.

Finally, we briefly discuss doping phase diagram for iron-based superconductors.
Figure \ref{Fig_Phs_dgrm} shows transition temperatures for SDW (determined by \msr) and $T_{\rm c}$ versus fluorine content $x$.
Since the entire sample is known to exhibit SDW phase for $x\le0.04$ in LFAOF \cite{Luetkens}, our result implies that LFAOF is an eutectic (``symbiotic") system of WM and SC domains over a narrow region $0.04<x<0.07$ (yellow-hatched area in Fig.~\ref{Fig_Phs_dgrm}).
This is in contrast to the case of another 1111 system CaFe$_{1-x}$Co$_x$AsF, in which the symbiotic region extends over a wide range of $0.05\le x\le0.15$ (Ref.\onlinecite{Takeshita_CFCAF}), while it is common to Ba(Fe$_{1-x}$Co$_x$)$_2$As$_2$ (Ref.\onlinecite{Bernhard, Williams}). These differences might come from that of Fermi surface structure and its response to carrier doping\cite{Vorontsov}.

In summary, our \msr\ experiment has shown occurrence of mesoscopic phase separation into superconducting, d-SDW and WM domains in \LFAOF\ with $x=0.057$ and the anomalous diamagnetism in the WM-domain. The diamagnetism suggests presence of a cooperative correlation between superconducting and magnetic order parameter in the WM phase, which is opposite to the competing order in regular SDW phase that microscopically coexists with superconductivity in Ba-122 family.

We would like to thank the staff of J-PARC MUSE and TRIUMF for their technical support during the $\mu$SR experiment, and F.~Esaka (JAEA) for SIMS measurement.

\begin{thebibliography}{00}
\bibitem{Kamihara} 
Y.~Kamihara, T.~Watanabe, M.~Hirano and H.~Hosono: \href{http://pubs.acs.org/doi/abs/10.1021/ja800073m}{J.~Am. Chem. Soc. {\bf 130}, (2008) 3296.}
\bibitem{Pratt} 
D.~K.~Pratt, W.~Tian, A.~Kreyssig, J.~L.~Zarestky, S.~Nandi, N.~Ni, S.~L.~Bud'ko, P.~C.~Canfield, A.~I.~Goldman and R.~J.~McQueeney: \href{http://prl.aps.org/abstract/PRL/v103/i8/e087001}{\PRL {\bf 103} (2009) 087001.}
\bibitem{Nandi} 
S.~Nandi, M.~G.~Kim, A.~Kreyssig, R.~M.~Fernandes, D.~K.~Pratt, A.~Thaler, N.~Ni, S.~L.~Bud'ko, P.~C.~Canfield, J.~Schmalian, R.~J.~McQueeney and A.~I. Goldman: \href{http://prl.aps.org/abstract/PRL/v104/i5/e057006}{\PRL {\bf 104} (2010) 057006.}
\bibitem{Aczel:08} A.~A.~Aczel, E. Baggio-Saitovitch, S. L. Budko, P. C. Canfield, J. P. Carlo, G. F. Chen, Pengcheng Dai, T. Goko, W. Z. Hu, G. M. Luke, J. L. Luo, N. Ni, D. R. Sanchez-Candela, F. F. Tafti, N. L. Wang, T. J. Williams, W. Yu, and Y. J. Uemura:
\href{http://prb.aps.org/abstract/PRB/v78/i21/e214503}{\PRB {\bf 78} (2008) 214503.}
\bibitem{Goko:09} T. Goko, A. A. Aczel, E. Baggio-Saitovitch, S. L. Bud¡Çko, P.~C. Canfield, J.~P. Carlo, G. F. Chen, Pengcheng Dai, A. C. Hamann, W. Z. Hu, H. Kageyama, G. M. Luke, J. L. Luo, B. Nachumi, N. Ni5, D. Reznik, D. R. Sanchez-Candela, A. T. Savici, K. J. Sikes, N. L. Wang, C. R. Wiebe, T. J. Williams, T. Yamamoto, W. Yu, and Y. J. Uemura,
\href{http://prb.aps.org/abstract/PRB/v80/i2/e024508}{\PRB {\bf 80} (2009) 024508.}
\bibitem{Park:09} J. T. Park, D. S. Inosov, Ch. Niedermayer, G. L. Sun, D. Haug, N. B. Christensen, R. Dinnebier, A. V. Boris, A. J. Drew, L. Schulz, T. Shapoval, U. Wolff, V. Neu, Xiaoping Yang, C. T. Lin, B. Keimer, and V. Hinkov,\href{http://prl.aps.org/abstract/PRL/v102/i11/e117006}{\PRL {\bf 102} (2009) 117006.}
\bibitem{Julien:09} M.-H. Julien, H. Mayaffre, M. Horvati?, C. Berthier, X. D. Zhang, W. Wu, G. F. Chen, N. L. Wang and J. L. Luo,
\href{http://iopscience.iop.org/0295-5075/87/3/37001/}{Europhys. Lett. {\bf 87} (2009) 37001.}
\bibitem{Wiesenmayer} 
E.~Wiesenmayer, H.~Luetkens, G.~Pascua, R.~Khasanov, A.~Amato, H.~Potts, B.~Banusch, H.-H.~Klauss and D.~Johrendt: \href{http://prl.aps.org/abstract/PRL/v107/i23/e237001}{\PRL {\bf 107} (2011) 237001.}
\bibitem{Takeshita_CFCAF} 
S.~Takeshita, R.~Kadono, M.~Hiraishi, M.~Miyazaki, A.~Koda, S.~Matsuishi and H.~Hosono: \href{http://prl.aps.org/abstract/PRL/v103/i2/e027002}{\PRL {\bf 103}, (2009) 027002.}
\bibitem{Takeshita_LFAOF} 
S.~Takeshita, R.~Kadono, M.~Hiraishi, M.~Miyazaki, A.~Koda, Y.~Kamihara and H.~Hosono: \href{http://jpsj.ipap.jp/link?JPSJ/77/103703/}{\JPSJ {\bf 77}, (2008) 103703.}
\bibitem{Prozorov} 
R.~Prozorov, M.~E.~Tillman, E.~D.~Mun and P.~C.~Canfield: \href{http://iopscience.iop.org/1367-2630/11/3/035004}{New~J. Phys. {\bf 11}, (2009) 035004.}
\bibitem{Zhigadlo} 
N.~D~Zhigadlo, S.~Katrych, Z.~Bukowski, S.~Weyeneth, R.~Puzniak and J.~Karpinski: \href{http://iopscience.iop.org/0953-8984/20/34/342202}{J.~Phys. Condens. Matter {\bf 20}, (2008) 342202.}
\bibitem{Jia} 
Y.~Jia, P.~Cheng, L.~Fang, H.~Luo, H.~Yang, C.~Ren, L.~Shan, C.~Gu and H.-H.~Wen: \href{http://apl.aip.org/resource/1/applab/v93/i3/p032503_s1}{Appl.~Phys. Lett. {\bf 93}, (2008) 032503.}
\bibitem{Ishikado_single} 
M.~Ishikado, S.~Shamoto, H.~Kito, A.~Iyo, H.~Eisaki, T.~Ito and Y.~Tomioka: \href{http://www.sciencedirect.com/science/article/pii/S0921453409001786}{Physica C {\bf 469}, (2009) 901.}
\bibitem{Ishikado} 
M.~Ishikado, R.~Kajimoto, S.~Shamoto, M.~Arai, A.~Iyo, K.~Miyazawa, P.~M.~Shirage, H.~Kito, H.~Eisaki, S.~Kim, H.~Hosono, T.~Guidi, R.~Bewley and S.~M.~Bennington: \href{http://jpsj.ipap.jp/link?JPSJ/78/043705/}{\JPSJ {\bf 78}, (2009) 043705.}
\bibitem{Wakimoto} 
S.~Wakimoto, K.~Kodama, M.~Ishikado, M.~Matsuda, R.~Kajimoto, M.~Arai, K.~Kakurai, F.~Esaka, A.~Iyo, H.~Kito, H.~Eisaki and S.~Shamoto: \href{http://jpsj.ipap.jp/link?JPSJ/79/074715/}{\JPSJ {\bf 79}, (2010) 074715.}
\bibitem{Kasahara:12}
See, for example, S. Kasahara, H. J. Shi, K. Hashimoto, S. Tonegawa, Y. Mizukami, T. Shibauchi, K. Sugimoto, T. Fukuda, T. Terashima, A. H. Nevidomskyy, and Y. Matsuda: \href{http://www.nature.com/nature/journal/v486/n7403/full/nature11178.html}{Nature {\bf 486}, (2012) 382}, and references therein.
\bibitem{Shahbazi} 
M.~Shahbazi, X.~L.~Wang, C.~Shekhar, O.~N.~Srivastava, Z.~W.~Lin, J.~G.~Zhu and S.~X.~Dou: \href{http://jap.aip.org/resource/1/japiau/v109/i7/p07E162_s1}{J.~Appl.~Phys. {\bf 109}, (2011) 07E162.}
\bibitem{Hunte} 
F.~Hunte, J.~Jaroszynski, A.~Gurevich, D.~C.~Larbalestier, R.~Jin, A.~S.~Sefat, M.~A.~McGuire, B.~C.~Sales, D.~K.~Christen and D.~Mandrus: \href{http://www.nature.com/nature/journal/v453/n7197/abs/nature07058.html}{Nature {\bf 453} (2008) 903.}
\bibitem{Iye} 
T.~Iye, Y.~Nakai, S.~Kitagawa, K.~Ishida, S.~Kasahara, T.~Shibauchi, Y.~Matsuda and T.~Terashima: \href{http://jpsj.ipap.jp/link?JPSJ/81/033701/}{\JPSJ {\bf 81} (2012) 033701.}
\bibitem{Fernandes:10}
R. M. Fernandes and J. Schmalian: \href{http://prb.aps.org/abstract/PRB/v82/i1/e014521}{\PRB {\bf 82} (2010) 014521.}
\bibitem{Qureshi:10}
N. Qureshi, Y. Drees, J. Werner, S. Wurmehl, C. Hess, R. Klingeler, B. B\"uchner, M. T. Fern\'andez-Diaz, and M. Braden: \href{http://prb.aps.org/abstract/PRB/v82/i18/e184521}{\PRB {\bf 82} (2010) 184521.}
\bibitem{Nomura:09}
T. Nomura, Y. Inoue, S. Matsuishi, M. Hirano, J.E. Kim, K. Kato, M. Takata and H. Hosono: \href{http://iopscience.iop.org/0953-2048/22/5/055016/}{Supercond. Sci.
Tech. {\bf 22} (2009) 055016.}
\bibitem{Koda:02} 
A. Koda, W. Higemoto, R. Kadono, Y. Kawasaki, K. Ishida, Y. Kitaoka, C. Geibel, and F. Steglich: \href{http://jpsj.ipap.jp/link?JPSJ/71/1427/}{\JPSJ {\bf 71}, 1427 (2002).} %
\bibitem{Luetkens} 
H.~Luetkens, H.-H.~Klauss, M.~Kraken, F.~J.~Litterst, T.~Dellmann, R.~Klingeler, C.~Hess, R.~Khasanov, A.~Amato, C.~Baines, M.~Kosmala, O.~J.~Schumann, M.~Braden, J.~Hamann-Borrero, N.~Leps, A.~Kondrat, G.~Behr, J.~Werner and B.~B$\ddot{\rm u}$chner: \href{http://www.nature.com/nmat/journal/v8/n4/full/nmat2397.html}{Nat.~Mater. {\bf 85}, (2009) 305.}
\bibitem{Carlo} 
J.~P.~Carlo, Y.~J.~Uemura, T.~Goko, G.~J.~MacDougall, J.~A.~Rodriguez, W.~Yu, G.~M.~Luke, Pengcheng~Dai, N. Shannon, S.~Miyasaka, S.~Suzuki, S.~Tajima, G.~F.~Chen, W.~Z.~Hu, J.~L.~Luo and N.~L.~Wang: \href{http://prl.aps.org/abstract/PRL/v102/i8/e087001}{\PRL {\bf 102}, (2009) 087001.}
\bibitem{Bernhard} 
C.~Bernhard, A.~J.~Drew, L.~Schulz, V.~K.~Malik, M.~R$\ddot{\rm o}$ssle, Ch~Niedermayer, Th.~Wolf, G.~D.~Varma, G.~Mu, H-H.~Wen, H.~Liu, G.~Wu and X.~H.~Chen: \href{http://iopscience.iop.org/1367-2630/11/5/055050}{New~J. Phys. {\bf 11}, (2009) 055050.}
\bibitem{Williams} 
T.~J.~Williams, A.~A.~Aczel, E.~Baggio-Saitovitch, S.~L.~Bud'ko, P.~C.~Canfield, J.~P.~Carlo, T.~Goko, J.~Munevar, N.~Ni, Y.~J.~Uemura, W.~Yu and G.~M.~Luke: \href{http://prb.aps.org/abstract/PRB/v80/i9/e094501}{\PRB {\bf 80}, (2009) 094501.}
\bibitem{Vorontsov} 
A.~B.~Vorontsov, M.~G.~Vavilov and A.~V.~Chubukov: \href{http://prb.aps.org/abstract/PRB/v81/i17/e174538}{\PRB {\bf 81}, (2010) 174538.}
\end{thebibliography}
\end{document}